\documentclass[aps,prl, twocolumn]{revtex4-1}

\pdfoutput=1

\usepackage{graphicx}
\usepackage{amsmath}
\usepackage{amsmath}
\usepackage{amsfonts}

\begin{document}

\title{Healing of Defects in Random Antiferromagnetic Spin Chains}

\author{R. Vasseur$^{1,2}$, A. Roshani$^{3}$, S. Haas$^{3}$ and  H. Saleur$^{3,4}$}

\affiliation{${}^1$Department of Physics, University of California, Berkeley, California 94720, USA}
\affiliation{${}^2$Materials Sciences Division, Lawrence Berkeley National Laboratory, Berkeley, California 94720, USA}
\affiliation{${}^3$Department of Physics and Astronomy,
University of Southern California, Los Angeles, CA 90089-0484, USA}
\affiliation{${}^4$Institut de Physique Th\'eorique, CEA Saclay,
91191 Gif Sur Yvette, France}

\date{\today}

\begin{abstract}
We study the effects of a weakened link in random antiferromagnetic spin chains. We show that {\sl healing} occurs, and that homogeneity is restored at low energy, in a way that is qualitatively similar to the fate of impurities in clean ferromagnetic spin chains, or in Luttinger liquids with attractive interactions. Healing in the random case occurs even without interactions, and is characteristic of the random singlet phase. Using real space renormalization group and exact diagonalization methods, we characterize this universal healing crossover by studying the entanglement across the weak link. We identify a crossover healing length $L^*$ that separates a regime where the system is cut in half by the weak link from a fixed point where the spin chain is healed. Our results open the way to the study of impurity physics in disordered spin chains.

\end{abstract}


\maketitle

\paragraph{Introduction.}

A potential scatterer in a Fermi liquid provides what is probably the simplest example of a quantum impurity problem. The associated physics has some interesting features -- it exhibits an Anderson orthogonality catastrophe~\cite{Anderson} and Friedel oscillations~\cite{Friedel} -- but remains trivial from the point of view of the Renormalization Group (RG):  properties are simply  determined by the phase shift at the Fermi-energy, and there is no crossover like, {\it e.g.}, in the Kondo problem~\cite{Hewson}.
 
The 1d version of the potential scatterer problem is easily realized in  a tight binding model with nearest neighbor hopping by modifying one hopping amplitude $J\to \lambda J$. The effect of the phase shift can then be seen e.g. in the zero temperature conductance, which is a simple function of the modified amplitude. 
As explained in the pioneering work of Ref.~\cite{KaneFisher}, interactions between the electrons profoundly change this picture.  When interactions are repulsive,  the electrons are completely reflected by even the smallest scatterer ($1-\lambda \ll 1$), and the $T=0$ linear conductance vanishes. In contrast, in the attractive case,   electrons are fully transmitted, even through a very strong scatterer ($\lambda \ll 1$), and the $T=0$ linear conductance takes the same value as for the homogenous system.  In both cases, the $I-V$ curves are non trivial and  exhibit various power law dependencies with non-trivial exponents. The situation has been well investigated both theoretically~\cite{Moonetal,FLS95} and experimentally~\cite{Glattli,Rezhnikov}. The physics can be understood in terms of the RG, and exhibits crossovers between the fixed point where the system is cut in half, and the fixed point where the system is ``healed''~\cite{Affleck}. Which of these fixed points is stable depends on the interaction. 
Instead of interacting electrons one can think equivalently of a spin chain. We restrict  for simplicity to the spinless case, which can be  described by an antiferromagnetic XXZ spin chain with anisotropy $\Delta$.  The region $0< \Delta \leq 1$ (resp. $-1<\Delta<0$)  corresponds to repulsive (resp. attractive) interactions.  Non interacting electrons correspond to $\Delta=0$, {\it i.e.} to the XX chain.

We discuss in this Letter the effect of introducing a modified link in a {\sl disordered} antiferromagnetic XXZ chain. More precisely, we consider the Hamiltonian
\begin{equation}
H=\sum_{i=-L-1}^{-1} J_i (\vec{S}_i.\vec{S}_{i+1})_\Delta+\lambda J_0 (\vec{S}_0.\vec{S}_1)_\Delta+\sum_{i=1}^{L-1} J_i(\vec{S}_i.\vec{S}_{i+1})_\Delta, \label{Hdis}
\end{equation}
where $\vec{S}$ is a spin $1/2$, and $(\vec{A}.\vec{B})_\Delta=A^xB^x+A^yB^y+\Delta A^zB^z$, the $J_i$ are independent random variables drawn from the same distribution, and $\lambda<1$ is a {\sl fixed parameter}. We shall show that, in contrast with the pure case, this system exhibits healing at low-energy for all values of the interaction $0\leq \Delta \leq 1$, including $\Delta=0$. Note that for a given realization of disorder, $\lambda J_0$ may be smaller or larger than most of the other $J_i's$. 
The distribution of the modified coupling $\lambda J_0$ is different from the distribution of all other couplings, and, if $\lambda<1$, favors smaller couplings: for simplicity we will keep referring  to the bond between sites $0$ and $1$ as the weak bond, and often denote it by B.

\paragraph{The random singlet phase.} The physics of model (\ref{Hdis}) without the modified coupling (ie, $\lambda=1$) is rather well understood as an infinite randomness RG fixed point. For essentially any probability distribution of the couplings $J_i$, it is known, based on exact solutions and real space renormalization group (RSRG) calculations, that the model flows into the  random singlet phase (RSP) with infinite disorder~\cite{FisherRSRG1,FisherRSRG2}. The corresponding fixed point theory is a quantum critical point  independent of $0\leq \Delta\leq 1$~\cite{FisherRSRG2}  which, while not conformally invariant, exhibits many properties similar to those of ordinary critical spin chains~\cite{MooreRefaelreview}: for instance, the entanglement entropy of a region of length $\ell$ with the rest of an infinite chain scales as $S \approx {\ln 2\over 3}\ln \ell$~\cite{MooreRefael1}. There has been a regain of interest in such infinite randomness critical points recently as they are also believed to describe dynamical (excited-state) phase transitions separating different many-body localized phases~\cite{PekkerRSRGX,VoskAltmanPRL13,QCGPRL}, and we note that our results  apply directly to some of these non-equilibrium transitions as well.

\paragraph{A weak link in the RSP.} Many properties could be used  to investigate  the behavior of the system (\ref{Hdis}) with $\lambda<1$. We will focus here  on entanglement  which, in contrast to the pure case, is  easier to calculate in the disordered case using the RSRG. For a chain with periodic boundary conditions and length $2L$, we consider  more precisely the entanglement of one half (of length $L$) with the other half.  In the pure case and when healing occurs, one expects that this entanglement at high energy should appear to be the entanglement of one half of an open chain with the other half (because the weak bond in that limit is essentially zero, and because we use periodic boundary conditions), thus $S^{\rm UV}\approx {c\over 6}\ln L$ with $c=1$ the central charge of the underlying conformal field theory~\cite{1742-5468-2004-06-P06002,1751-8121-42-50-504005}. Meanwhile, at low-energy, the system should appear healed, and one should recover the entanglement of half a homogeneous periodic chain with the other half, $S^{\rm IR}\approx {c\over 3}\ln L$. In between,  the entanglement interpolates between these two behaviors, and, in the limit of small $\lambda$ and large $L$, it can be shown that the function $\partial S/\partial\ln L$ is a {\sl universal function} over $L/L^\star$, where the crossover length $L^\star$ is a power of the weak coupling, similar to the Kondo length $\xi_K$ in Kondo problems~\cite{TbRef,VJS14,EEIRLM}. 

We shall see that a similar behavior occurs in the RSP. The entanglement extrapolates now between $S^{\rm UV}\approx {\ln 2\over 6}\ln L$ and $S^{\rm IR}\approx{\ln 2\over 3}\ln L$. Moreover, in the limit of small $\lambda$ and large $L$, $\partial S/ \partial\ln L$ is a universal function of $L/L^*$, where now the crossover length is $L^*\propto (-\ln \lambda)^2$. Hence, despite the presence of disorder -  which could be feared to erase all interesting impurity physics -  healing in fact does occur. In contrast with the pure case, healing occurs in the antiferromagnetic (repulsive) case, including the free case, and the scaling function is independent of the anisotropy $\Delta$. 

\paragraph{Healing in the RSP.} 
We now describe the physical idea behind the result, together with the essential steps of the calculation. Details are given in the supplemental material~\cite{SupMat}. Recall that in the RSP, the ground state of the system is almost factorized into pairs of singlets at arbitrarily large distances.  Properties are best studied using the RSRG. In this approach, one starts with the strongest bond, which is projected onto  a singlet, while  the neighboring spins  inherit, from second-order perturbation theory, an effective interaction~\cite{PhysRevLett.43.1434,PhysRevB.22.1305}. This ``decimation'' corresponds, in terms of couplings, to  $(\ldots,J_{i-1},J_i,J_{i+1},\ldots)_{2L}\to (\ldots, J_{i-1}J_{i+1}/J_i,\ldots)_{2L-2}$ where $J_i$ is the strongest bond, $J_{i-1},J_{i+1}$ the  coupling on the immediate left or right. One can then iterate this procedure, and it is known that, if one starts with the same distribution for all bonds ({\it i.e.}, $\lambda=1$), after a sufficiently large number of decimations, the distribution of couplings $\beta_i\equiv \ln{\Omega\over J_i}$ (with $\Omega = {\rm max} \lbrace J_i \rbrace$ the reduced energy scale obtained after iterating the RG) converges, independently of the original distribution, to the universal form
\begin{equation}
P_\Gamma(\beta)={1\over \Gamma}e^{-\beta/\Gamma}, \label{FPdist}
\end{equation}
where $\Gamma$ is the RG flow parameter, $\Gamma=\ln (\Omega_0/\Omega)$, and $\Omega_0$  is the initial energy scale (UV cutoff).

This RG picture is particularly adapted to study the entanglement, since, in a given RSP configuration, the entanglement of a region with its complement is simply proportional to the number of singlets $n$ connecting the two~\cite{MooreRefael1,MooreRefaelreview}: $S=n \ln 2$. In the presence of the impurity and for $\lambda$ small enough,  the weak bond will not, most of the time be decimated in the early stages of the RG. In this case, no singlet can connect the two halves of the system across the weak bond, and their mutual entanglement will be the same as if the chain were cut at the origin. On the other hand, when the RG has been run long enough for even this weak bond to be decimated, the entanglement should be the same as with a chain with homogeneous disorder. Hence, some sort of ``healing'' is well expected to take place. Moreover, we expect the crossover energy in the RG to correspond to  $\Gamma \sim -\ln \lambda$. Like in the pure case, we will ultimately wish to replace the energy scale in the RG by a {\sl length scale}, {\it i.e.}, study the entanglement of the two halves of the chain as a function of the length $L$. The crossover scale in energy then should  translate into a crossover length scale $L^\star\sim (-\ln\lambda)^2$ using the  characteristic scaling $\ln [\hbox{energy}]\propto [\hbox{length}]^{\frac{1}{2}}$ of the RSP. 

\paragraph{RG analysis of the weak link.} It is clear from this picture that an essential quantity to characterize the crossover and the entanglement is the {\sl probability $P_0$ for the weak link not to be decimated during the RG}.  Since we are interested in universal quantities, we will assume that the initial distribution of couplings is already at the fixed point~(\ref{FPdist}). The probability distribution for the weak link is initially
\begin{equation}
Q_{\Gamma_0} (\beta) = \frac{1}{\Gamma_0} {\rm e}^{-\frac{\beta + \ln \lambda}{\Gamma_0}} \theta(\beta + \ln \lambda),
\label{InCond}
\end{equation}
with $\Gamma_0$ the initial disorder strength, since initially the weak link has $\beta = \ln \frac{\Omega_0}{\lambda J_0} >  \ln \lambda^{-1}$. Clearly, the weak bond cannot be decimated until the scale $\Gamma - \Gamma_0 = - \ln \lambda$. We now consider the probability $p_\Gamma$ that  the weak bond has not yet been  decimated at scale $\Gamma (> \Gamma_0)$.  Running the RG will involve decimating neighbors of B that will renormalize the distribution of couplings $Q_\Gamma(\beta)$, and following Ref.~\cite{MooreRefael1}, we will use the normalization $\int_{0}^\infty Q_{\Gamma}(\beta) d \beta = p_\Gamma$. The distribution $Q_{\Gamma}(\beta)$ then satisfies the flow equation~\cite{MooreRefael1} 
\begin{equation}
\frac{\partial Q_\Gamma}{\partial \Gamma} = \frac{\partial Q_\Gamma}{\partial \beta} + 2 P_\Gamma(0) \left(P_\Gamma \star Q_\Gamma - Q_\Gamma \right), \label{floweq}
\end{equation}
and we have $\frac{d p_\Gamma}{d \Gamma} = - Q_{\Gamma}(0)$. Computing $p_\Gamma$ therefore amounts to solving this partial differential equation with the initial condition~\eqref{InCond}. We solve this equation by introducting the Laplace transform $\hat{Q}_\Gamma(s) = \int_0^\infty d \beta {\rm e}^{-\beta s} Q_{\Gamma}(\beta)$: $\hat{Q}_\Gamma(s)$ can then be expressed in terms of the unknown probability $p_\Gamma$, and inverting the Laplace transform using the condition $\frac{d p_\Gamma}{d \Gamma} = - Q_{\Gamma}(0)$ yields an integral equation for $p_\Gamma$~\cite{SupMat}. More elegantly, we notice the probability of decimating the weak link is exactly zero ($p_\Gamma=1$) until $\Gamma$ reaches the crossover scale $\Gamma<\Gamma^\star \equiv \Gamma_0 - \ln \lambda$. After a straightforward calculation, we find that for $\Gamma<\Gamma^\star$ 
\begin{equation}
\hat{Q}_\Gamma (s) = \left( \frac{1+s \Gamma_0}{1+s \Gamma} \right)^2 {\rm e}^{s(\Gamma - \Gamma_0)} \frac{\lambda^s}{1+\Gamma_0 s}.
\end{equation}
We now consider the scaling limit $\Gamma \to \infty$, $\lambda \to 0$ with $x = \frac{\Gamma}{-\ln \lambda}$ fixed: this amounts to sending $\Gamma_0 \to 0$. Inverting the Laplace transform, we find that right at the crossover scale $\Gamma = \Gamma^\star$, the distribution
\begin{equation}
Q_{\Gamma\equiv -\ln\lambda}(\beta)={\beta\over (\ln\lambda)^2}e^{\beta/\ln\lambda},
\label{eqQgammastar}
\end{equation}
coincides with the distribution of a bond that has just been decimated $\int d \beta_L d \beta_R \delta(\beta - \beta_L - \beta_R) P_{\Gamma^\star} (\beta_R) P_{\Gamma^\star} (\beta_L)$. Using the results of Ref.~\cite{MooreRefael1}, we can then solve the flow equation~\eqref{floweq} for $\Gamma> \Gamma^\star$ using this effective ``initial condition''~\eqref{eqQgammastar} at $\Gamma=\Gamma^\star$. In the scaling limit, the function $p_\Gamma$ becomes a universal function $p(x)$ and we have  
\begin{equation}
p(x) = \frac{1 - \sqrt{5}}{5+\sqrt{5}} x^{- \frac{3+\sqrt{5}}{2}} + \frac{4 + 2 \sqrt{5}}{5+\sqrt{5}}  x^{- \frac{3-\sqrt{5}}{2}}, \ x>1,
\label{eqPPx}
\end{equation}
while  $p(x)=1$ for $x \leq 1$ (the weak link is not decimated until the crossover scale).

\begin{figure}[t!]
\includegraphics[width=1.0\linewidth]{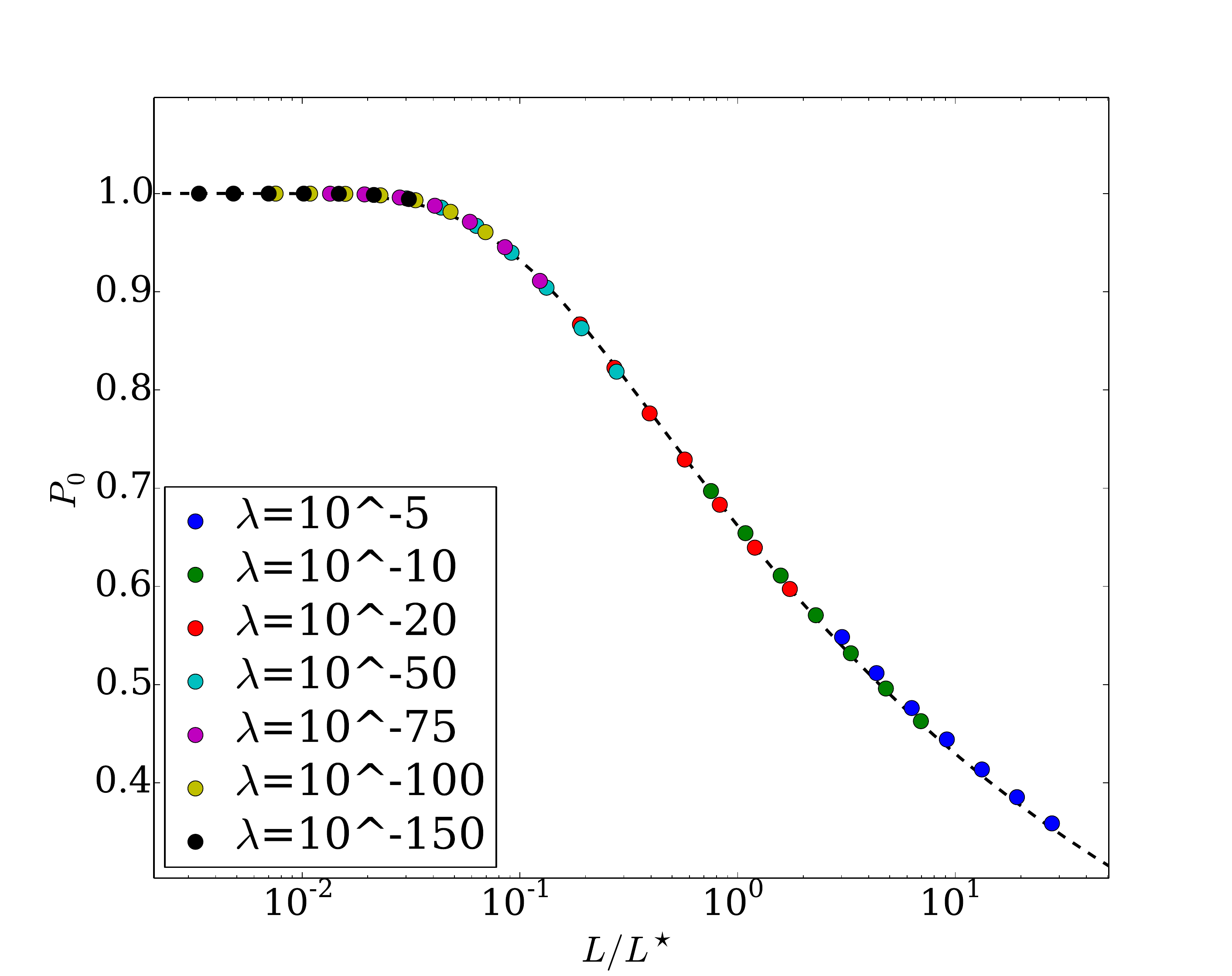}
\caption{Probability $P_0$ for the weak link not to be decimated (corresponding to vanishing entanglement across the link) for a system of size $2L$. The exact solution~\eqref{eqP0} (dashed line) agrees well with numerical RG calculations for various system sizes $2L \sim 400 - 4000$, with crossover scale $L^\star \sim (\ln \lambda)^2$.}
\label{figP0}
\end{figure}

Note that $p$ depends on the RG scale $\Gamma$.  We still need to trade it for the length scale in our problem. To do this we observe that, since we are dealing with a system of finite size, the RG will stop once all spins have been paired into singlets.We thus need another  ingredient: the distribution of the strength $\Omega$ (or $\Gamma=-\ln\Omega$) of the last bond being decimated. This is known as the  distribution of the first gap, and  was computed exactly in \cite{FisherYoung} as   $d P_{\rm gap} = \frac{1}{\Gamma} g(y=\Gamma/\sqrt{L}) d \Gamma$ with $g(y) = \frac{2 \pi}{y^2} \sum_{n=0}^{\infty} (-1)^n (2 n +1) {\rm e}^{-(2 n + 1)^2 \pi^2/4y^2}$. Gathering these different pieces, we can finally compute the quantity $P_0$ as 
\begin{equation}
P_0 = \int_0^\infty d \Gamma p\left(\frac{\Gamma}{-\ln \lambda} \right)  \frac{1}{\Gamma} g \left( \frac{\Gamma}{\sqrt{L}} \right), \label{eqP0}
\end{equation}
which is clearly a function of $L/L^\star$ with $L^\star = (-\ln \lambda)^2$. The explicit expression can be worked out in terms of generalized Gamma functions.
 %
%
Physically, this quantity controls the probability to have vanishing entanglement $S \leq \epsilon$ across the weak link for a system of size $2L$ with $\epsilon \ll 1$ arbitrarily small. We have $P_0(0)=1$ and $P_0 \sim L^{-(3-\sqrt{5})/4}$ for $L \gg L^\star$. The scaling function~\eqref{eqP0} agrees well with numerical RG calculations (Fig.~\ref{figP0}). 

\paragraph{Entanglement across the weak link.}

\begin{figure}[t!]
\includegraphics[width=1.0\linewidth]{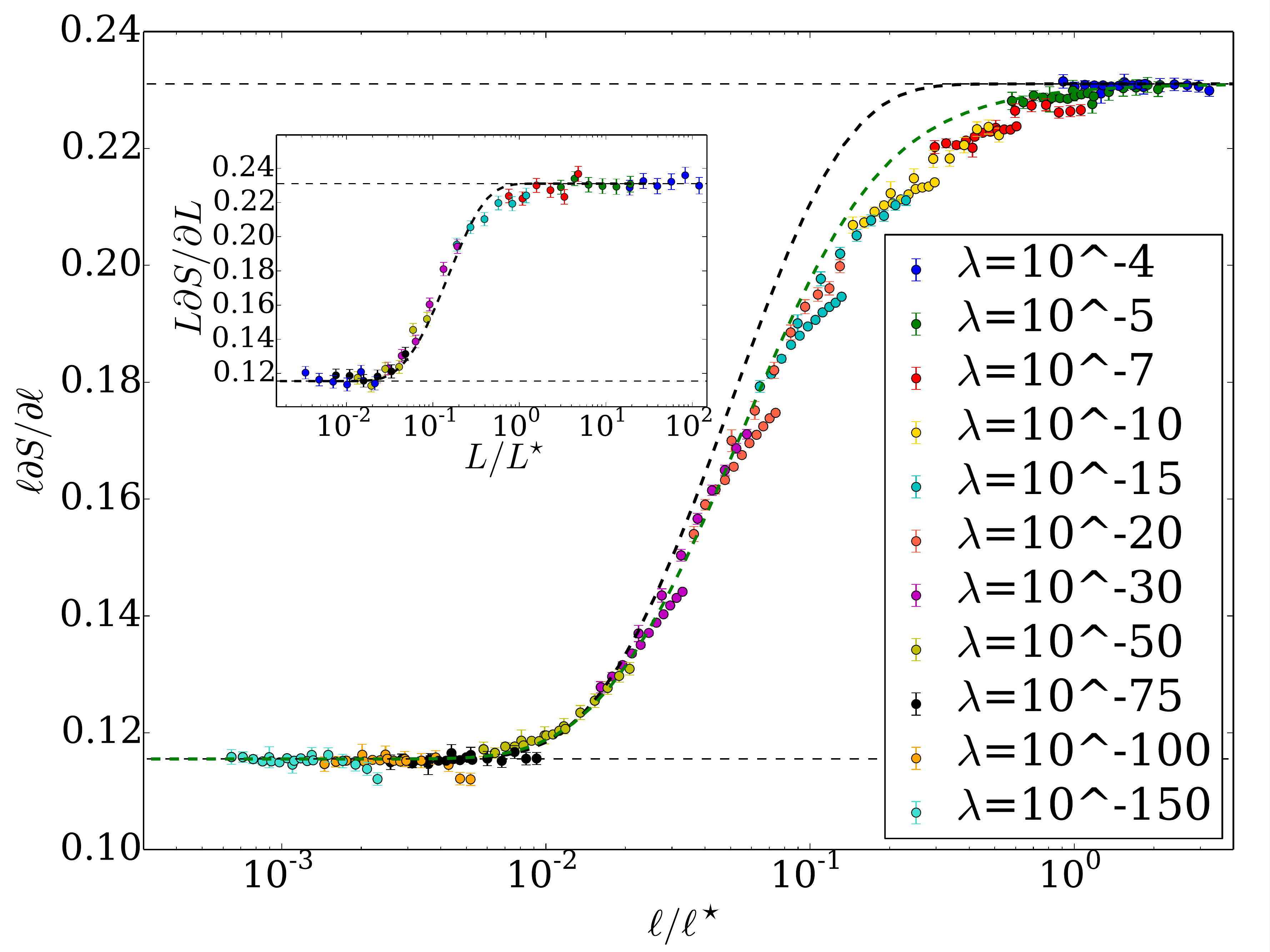}
\caption{Numerical RG calculation of the entanglement entropy of an interval of size $\ell \ll 2 L\sim 4000$ with a weak link of strength $\lambda$ at one extremity, inducing a crossover scale $\ell^\star \sim (\ln \lambda)^2$. The black dashed curve corresponds to eq.~\eqref{eqScrossover} interpolating between $\frac{\ln 2}{6}$ and $\frac{\ln2}{3}$, while the green dashed curve is including corrections to that formula (see text). Inset: similar crossover for the bipartite entanglement of a periodic system of size $2L$ showing various values of $\lambda$.
}
\label{fig2}
\end{figure}

Let us now go back to the entanglement calculation. Following~\cite{MooreRefael1}, the average ``RG time'' corresponding to the decimation of the weak link is $\ell \equiv \overline{\ln \frac{\Gamma}{-\ln \lambda}} = \int_0^\infty dx \frac{dp}{dx} \ln x = 3$, where we used eq.~\eqref{eqPPx}. This is the same value that Refael and Moore find for the average RG time separating decimations increasing the entanglement entropy by $\ln 2$.
 This is once again consistent with the very simple picture described above: with or without the weak link, the average RG time between decimations increasing the entanglement entropy by $\ln 2$ is given by $\ell = 3$, but the effect of the weak link is merely to shift the origin of time, since the first decimation typically occurs when $\overline{\ln \frac{\Gamma_{\rm typ}}{-\ln \lambda}} =3$, with in particular $\Gamma_{\rm typ}> - \ln \lambda$. This gives for the bipartite entanglement entropy 
\begin{equation}
S = \int_{\Gamma_0}^\infty    \frac{d \Gamma}{\Gamma} g \left( \frac{\Gamma}{\sqrt{L}} \right) \ln 2 \left[ \int_{\Gamma_0}^\Gamma  \frac{d \Gamma^\prime/\Gamma^\prime}{3} (1+ \theta(\Gamma^\prime + \ln \lambda))\right],
\end{equation}
up to non-universal contributions. Here, the first term arises from the singlets connecting the two halves over the periodic boundary condition, the second term arises from the weak link.  Going to the scaling limit, and considering the derivative with respect to $\ln L$ to get rid of the non-universal terms, we find~\cite{SupMat}
\begin{equation}
L \frac{\partial S}{ \partial L} = \frac{\ln 2}{3} \left( 1  - \frac{2}{\pi} \sum_{n=0}^\infty \frac{(-1)^n}{2n+1} {\rm e}^{-\pi^2 (2n+1)^2 L /4 L^\star} \right).
\label{eqScrossover}
\end{equation}
We remark that even though we considered the bipartite entanglement of a periodic system of size $2L$, eq.~\eqref{eqScrossover} with $L \to \ell$ should also describe the scaling of the entanglement of an interval $[0,\ell]$ with the weak link at $x=0$ in a system of size $L \gg \ell \gg 1$ (Fig.~\ref{fig2}). Note also that contrary to eq.~\eqref{eqP0}, we do not expect the scaling function~\eqref{eqScrossover} to be exact. This is because in deriving~\eqref{eqScrossover}, we used the fact that the number of singlets crossing a generic link in the system scales as $n \sim \frac{1}{3} \ln \frac{\Gamma}{\Gamma0}$ in terms of the RG scale $\Gamma$~\cite{MooreRefael1}. However, there exist universal power-law corrections to that formula~\cite{CalabreseMoore,SupMat} that are usually unimportant but that turn out to be crucial to compute the scaling function~\eqref{eqScrossover} exactly. Taking into account these corrections does not change fundamentally the qualitative behavior of eq.~\eqref{eqScrossover} but it does improve the quantitative agreement with numerical RG calculations (see Fig.~\ref{fig2}). Following Refs.~\cite{CalabreseMoore,2016arXiv161202012D}, the arguments above can easily be generalized to the calculate  the full probability distribution of entanglement across the weak link~\cite{SupMat} (or equivalently, the entanglement spectrum).

\paragraph{Numerical calculations.}

Even though the RSRG approach is clearly not exact for finite chains, it is expected to give an asymptotically exact description of universal properties~\cite{FisherRSRG1,FisherRSRG2}. We have attempted to study the scaling regime $L\to\infty,\lambda\to 0$ numerically by directly determining the entanglement of a random XX chain with a weak link. The approach is well known~\cite{0305-4470-38-20-002,Peschel}, and discussed in detail in Ref.~\cite{Laflorencie}. Formulating the chain as a model of random hopping fermions, the reduced density matrix is determined by the eigenvalues of the correlation matrix $\langle c_i^\dagger c_j\rangle$, where the average is to be calculated in the ground state of the disordered system, and the labels $i,j$ run from $1$ to $L$. 
Because of numerical instabilities for small $\lambda $ and large $L$, we have only been able to obtain partial scaling collapses, which are however in good agreement with  eq.~\eqref{eqScrossover}. By fitting our numerical results for the entanglement entropy to~\eqref{eqScrossover} for different values of $\lambda$ to extract the optimal healing scale $L^\star (\lambda)$, we were able also to verify the scaling  $L^\star \sim (\ln \lambda)^2$. This crossover scale is remarkably short: even for a very weak link $\lambda \sim 10^{-10}$, the system appears ``healed''  (so that the impurity or weak link can essentially be ignored) on a length scale of order $10^2 - 10^3$ sites.

\begin{figure}[t!]
\includegraphics[width=1.0\linewidth]{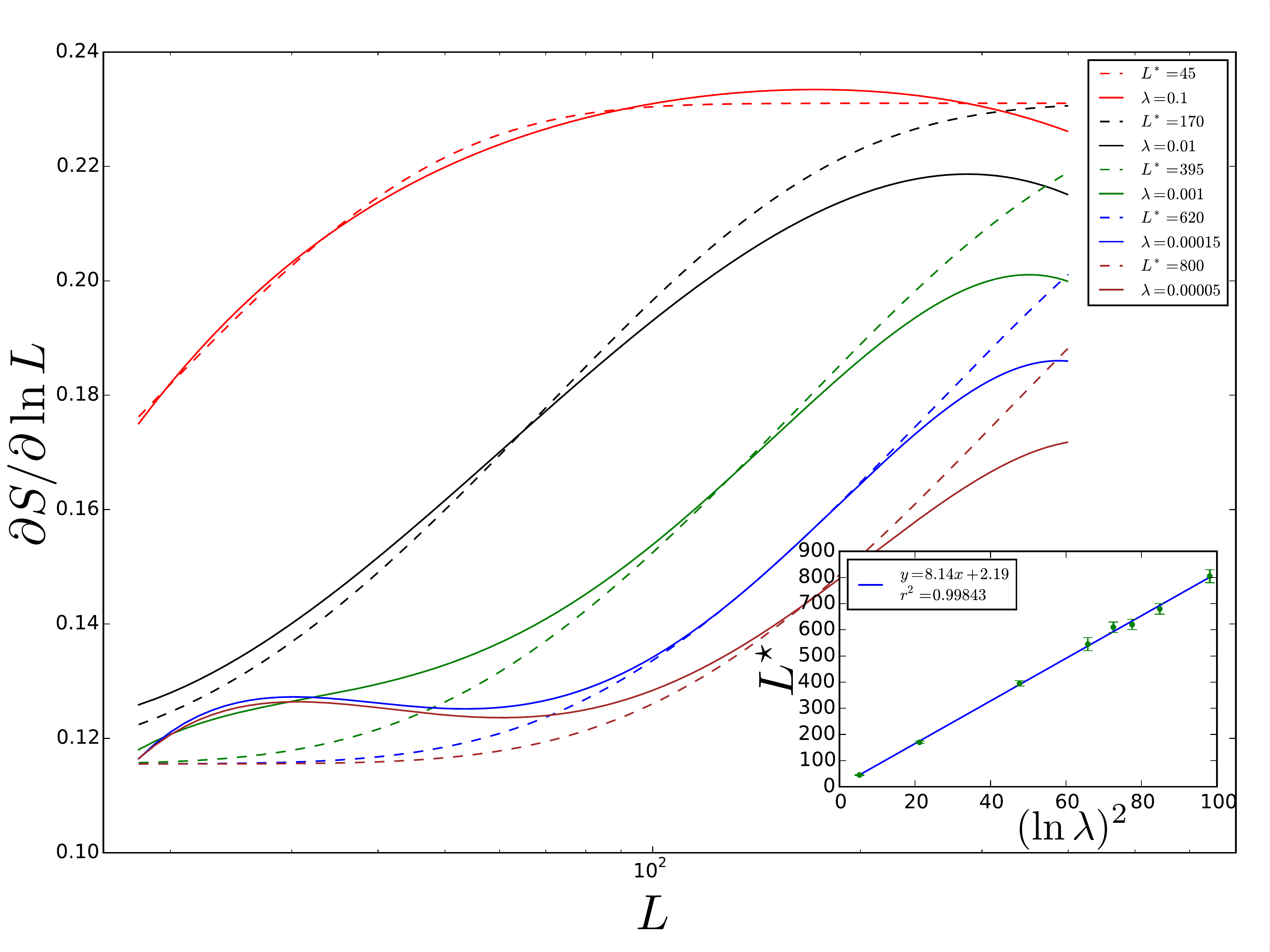}
\caption{Numerical determination of the entanglement in the random XX chain obtained by averaging over large numbers of realizations with various choices $L$ and $\lambda$, and  couplings randomly chosen in the interval $[0,1]$. The main frame shows numerical results  for the slope $\partial S/\partial \ln L$ against $L$ (full lines) versus the theoretical prediction~(\ref{eqScrossover}) after the best $L^*$ for this value of $\lambda$ has been determined. Inset: fitted $L^*$ against $(\ln \lambda)^2$.}
\label{fignumerics}
\end{figure}

\paragraph{Conclusion.}
 
 Remarkably, we have found that, in the case of disordered antiferromagnetic chains, healing occurs even without interactions, and is associated with a rich crossover physics. The existence of this healing flow is compatible with the results of Ref.~\cite{MooreVasseur} where it was found that the ground state of the chain with $\lambda=1$ and the chain with $0<\lambda<1$ are not orthogonal, and that there is no Anderson catastrophe. The same would be observed in a clean XXZ chain in the attractive regime. Since the repulsive or attractive nature of the interactions is so crucial for determining the fate of weak links in clean Luttinger liquids, it would be interesting to investigate the effect of weak perturbations in random ferromagnetic chains~\cite{PhysRevLett.75.4302,Monthus} to determine whether they also ``heal'' at low energy. 
  
 Although we were not able to find an exact scaling function for the entanglement, the RSRG does give   analytical and numerical access to the crossover. This should open the way to studying on the one hand more physical quantities of interest (such as t transport properties across the weak link), or to studying richer  setups. The simplest one in this case would be a problem with two weak links, where, in the UV, one should see a decoupled two-state system, reminiscent of a Kondo impurity. Indeed, it is known in the pure case that the resonant level model and its interacting variants are deeply related to the (anisotropic) Kondo model~\cite{FTW80}. We expect some interesting generalizations of this in the disordered case. 
 
\paragraph{Acknowledgments.} R.V. thanks Joel Moore for insightful discussions and for collaborations on related matters.
 This work was supported by the US Department of Energy grant number DE-FG03-01ER45908 (S.H. and H.S.), the Advanced ERC Grant NUQFT (H.S.) and the LDRD program of LBNL (R.V.).

 \bibliography{Disorder.bib}

\begin{thebibliography}{34}%
\makeatletter
\providecommand \@ifxundefined [1]{%
 \@ifx{#1\undefined}
}%
\providecommand \@ifnum [1]{%
 \ifnum #1\expandafter \@firstoftwo
 \else \expandafter \@secondoftwo
 \fi
}%
\providecommand \@ifx [1]{%
 \ifx #1\expandafter \@firstoftwo
 \else \expandafter \@secondoftwo
 \fi
}%
\providecommand \natexlab [1]{#1}%
\providecommand \enquote  [1]{``#1''}%
\providecommand \bibnamefont  [1]{#1}%
\providecommand \bibfnamefont [1]{#1}%
\providecommand \citenamefont [1]{#1}%
\providecommand \href@noop [0]{\@secondoftwo}%
\providecommand \href [0]{\begingroup \@sanitize@url \@href}%
\providecommand \@href[1]{\@@startlink{#1}\@@href}%
\providecommand \@@href[1]{\endgroup#1\@@endlink}%
\providecommand \@sanitize@url [0]{\catcode `\\12\catcode `\$12\catcode
  `\&12\catcode `\#12\catcode `\^12\catcode `\_12\catcode `\%12\relax}%
\providecommand \@@startlink[1]{}%
\providecommand \@@endlink[0]{}%
\providecommand \url  [0]{\begingroup\@sanitize@url \@url }%
\providecommand \@url [1]{\endgroup\@href {#1}{\urlprefix }}%
\providecommand \urlprefix  [0]{URL }%
\providecommand \Eprint [0]{\href }%
\providecommand \doibase [0]{http://dx.doi.org/}%
\providecommand \selectlanguage [0]{\@gobble}%
\providecommand \bibinfo  [0]{\@secondoftwo}%
\providecommand \bibfield  [0]{\@secondoftwo}%
\providecommand \translation [1]{[#1]}%
\providecommand \BibitemOpen [0]{}%
\providecommand \bibitemStop [0]{}%
\providecommand \bibitemNoStop [0]{.\EOS\space}%
\providecommand \EOS [0]{\spacefactor3000\relax}%
\providecommand \BibitemShut  [1]{\csname bibitem#1\endcsname}%
\let\auto@bib@innerbib\@empty
\bibitem [{\citenamefont {Anderson}(1967)}]{Anderson}%
  \BibitemOpen
  \bibfield  {author} {\bibinfo {author} {\bibfnamefont {P.~W.}\ \bibnamefont
  {Anderson}},\ }\href {\doibase 10.1103/PhysRevLett.18.1049} {\bibfield
  {journal} {\bibinfo  {journal} {Phys. Rev. Lett.}\ }\textbf {\bibinfo
  {volume} {18}},\ \bibinfo {pages} {1049} (\bibinfo {year}
  {1967})}\BibitemShut {NoStop}%
\bibitem [{\citenamefont {Friedel}(1952)}]{Friedel}%
  \BibitemOpen
  \bibfield  {author} {\bibinfo {author} {\bibfnamefont {J.}~\bibnamefont
  {Friedel}},\ }\href {\doibase 10.1080/14786440208561086} {\bibfield
  {journal} {\bibinfo  {journal} {The London, Edinburgh, and Dublin
  Philosophical Magazine and Journal of Science}\ }\textbf {\bibinfo {volume}
  {43}},\ \bibinfo {pages} {153} (\bibinfo {year} {1952})}\BibitemShut
  {NoStop}%
\bibitem [{\citenamefont {Hewson}(1997)}]{Hewson}%
  \BibitemOpen
  \bibfield  {author} {\bibinfo {author} {\bibfnamefont {A.~C.}\ \bibnamefont
  {Hewson}},\ }\href@noop {} {\emph {\bibinfo {title} {The Kondo problem to
  heavy fermions}}},\ Vol.~\bibinfo {volume} {2}\ (\bibinfo  {publisher}
  {Cambridge university press},\ \bibinfo {year} {1997})\BibitemShut {NoStop}%
\bibitem [{\citenamefont {Kane}\ and\ \citenamefont
  {Fisher}(1992)}]{KaneFisher}%
  \BibitemOpen
  \bibfield  {author} {\bibinfo {author} {\bibfnamefont {C.~L.}\ \bibnamefont
  {Kane}}\ and\ \bibinfo {author} {\bibfnamefont {M.~P.~A.}\ \bibnamefont
  {Fisher}},\ }\href {\doibase 10.1103/PhysRevLett.68.1220} {\bibfield
  {journal} {\bibinfo  {journal} {Phys. Rev. Lett.}\ }\textbf {\bibinfo
  {volume} {68}},\ \bibinfo {pages} {1220} (\bibinfo {year}
  {1992})}\BibitemShut {NoStop}%
\bibitem [{\citenamefont {Moon}\ \emph {et~al.}(1993)\citenamefont {Moon},
  \citenamefont {Yi}, \citenamefont {Kane}, \citenamefont {Girvin},\ and\
  \citenamefont {Fisher}}]{Moonetal}%
  \BibitemOpen
  \bibfield  {author} {\bibinfo {author} {\bibfnamefont {K.}~\bibnamefont
  {Moon}}, \bibinfo {author} {\bibfnamefont {H.}~\bibnamefont {Yi}}, \bibinfo
  {author} {\bibfnamefont {C.~L.}\ \bibnamefont {Kane}}, \bibinfo {author}
  {\bibfnamefont {S.~M.}\ \bibnamefont {Girvin}}, \ and\ \bibinfo {author}
  {\bibfnamefont {M.~P.~A.}\ \bibnamefont {Fisher}},\ }\href {\doibase
  10.1103/PhysRevLett.71.4381} {\bibfield  {journal} {\bibinfo  {journal}
  {Phys. Rev. Lett.}\ }\textbf {\bibinfo {volume} {71}},\ \bibinfo {pages}
  {4381} (\bibinfo {year} {1993})}\BibitemShut {NoStop}%
\bibitem [{\citenamefont {Fendley}\ \emph {et~al.}(1995)\citenamefont
  {Fendley}, \citenamefont {Ludwig},\ and\ \citenamefont {Saleur}}]{FLS95}%
  \BibitemOpen
  \bibfield  {author} {\bibinfo {author} {\bibfnamefont {P.}~\bibnamefont
  {Fendley}}, \bibinfo {author} {\bibfnamefont {A.~W.~W.}\ \bibnamefont
  {Ludwig}}, \ and\ \bibinfo {author} {\bibfnamefont {H.}~\bibnamefont
  {Saleur}},\ }\href {\doibase 10.1103/PhysRevLett.74.3005} {\bibfield
  {journal} {\bibinfo  {journal} {Phys. Rev. Lett.}\ }\textbf {\bibinfo
  {volume} {74}},\ \bibinfo {pages} {3005} (\bibinfo {year}
  {1995})}\BibitemShut {NoStop}%
\bibitem [{\citenamefont {Saminadayar}\ \emph {et~al.}(1997)\citenamefont
  {Saminadayar}, \citenamefont {Glattli}, \citenamefont {Jin},\ and\
  \citenamefont {Etienne}}]{Glattli}%
  \BibitemOpen
  \bibfield  {author} {\bibinfo {author} {\bibfnamefont {L.}~\bibnamefont
  {Saminadayar}}, \bibinfo {author} {\bibfnamefont {D.~C.}\ \bibnamefont
  {Glattli}}, \bibinfo {author} {\bibfnamefont {Y.}~\bibnamefont {Jin}}, \ and\
  \bibinfo {author} {\bibfnamefont {B.}~\bibnamefont {Etienne}},\ }\href
  {\doibase 10.1103/PhysRevLett.79.2526} {\bibfield  {journal} {\bibinfo
  {journal} {Phys. Rev. Lett.}\ }\textbf {\bibinfo {volume} {79}},\ \bibinfo
  {pages} {2526} (\bibinfo {year} {1997})}\BibitemShut {NoStop}%
\bibitem [{\citenamefont {de~Picciotto}\ \emph {et~al.}(1997)\citenamefont
  {de~Picciotto}, \citenamefont {Reznikov}, \citenamefont {Heiblum},
  \citenamefont {Umansky}, \citenamefont {Bunin},\ and\ \citenamefont
  {Mahalu}}]{Rezhnikov}%
  \BibitemOpen
  \bibfield  {author} {\bibinfo {author} {\bibfnamefont {R.}~\bibnamefont
  {de~Picciotto}}, \bibinfo {author} {\bibfnamefont {M.}~\bibnamefont
  {Reznikov}}, \bibinfo {author} {\bibfnamefont {M.}~\bibnamefont {Heiblum}},
  \bibinfo {author} {\bibfnamefont {V.}~\bibnamefont {Umansky}}, \bibinfo
  {author} {\bibfnamefont {G.}~\bibnamefont {Bunin}}, \ and\ \bibinfo {author}
  {\bibfnamefont {D.}~\bibnamefont {Mahalu}},\ }\href@noop {} {\bibfield
  {journal} {\bibinfo  {journal} {Nature}\ }\textbf {\bibinfo {volume} {389}},\
  \bibinfo {pages} {162} (\bibinfo {year} {1997})}\BibitemShut {NoStop}%
\bibitem [{\citenamefont {Eggert}\ and\ \citenamefont
  {Affleck}(1992)}]{Affleck}%
  \BibitemOpen
  \bibfield  {author} {\bibinfo {author} {\bibfnamefont {S.}~\bibnamefont
  {Eggert}}\ and\ \bibinfo {author} {\bibfnamefont {I.}~\bibnamefont
  {Affleck}},\ }\href {\doibase 10.1103/PhysRevB.46.10866} {\bibfield
  {journal} {\bibinfo  {journal} {Phys. Rev. B}\ }\textbf {\bibinfo {volume}
  {46}},\ \bibinfo {pages} {10866} (\bibinfo {year} {1992})}\BibitemShut
  {NoStop}%
\bibitem [{\citenamefont {Fisher}(1992)}]{FisherRSRG1}%
  \BibitemOpen
  \bibfield  {author} {\bibinfo {author} {\bibfnamefont {D.~S.}\ \bibnamefont
  {Fisher}},\ }\href {\doibase 10.1103/PhysRevLett.69.534} {\bibfield
  {journal} {\bibinfo  {journal} {Phys. Rev. Lett.}\ }\textbf {\bibinfo
  {volume} {69}},\ \bibinfo {pages} {534} (\bibinfo {year} {1992})}\BibitemShut
  {NoStop}%
\bibitem [{\citenamefont {Fisher}(1994)}]{FisherRSRG2}%
  \BibitemOpen
  \bibfield  {author} {\bibinfo {author} {\bibfnamefont {D.~S.}\ \bibnamefont
  {Fisher}},\ }\href {\doibase 10.1103/PhysRevB.50.3799} {\bibfield  {journal}
  {\bibinfo  {journal} {Phys. Rev. B}\ }\textbf {\bibinfo {volume} {50}},\
  \bibinfo {pages} {3799} (\bibinfo {year} {1994})}\BibitemShut {NoStop}%
\bibitem [{\citenamefont {Refael}\ and\ \citenamefont
  {Moore}(2009)}]{MooreRefaelreview}%
  \BibitemOpen
  \bibfield  {author} {\bibinfo {author} {\bibfnamefont {G.}~\bibnamefont
  {Refael}}\ and\ \bibinfo {author} {\bibfnamefont {J.~E.}\ \bibnamefont
  {Moore}},\ }\href {http://stacks.iop.org/1751-8121/42/i=50/a=504010}
  {\bibfield  {journal} {\bibinfo  {journal} {Journal of Physics A:
  Mathematical and Theoretical}\ }\textbf {\bibinfo {volume} {42}},\ \bibinfo
  {pages} {504010} (\bibinfo {year} {2009})}\BibitemShut {NoStop}%
\bibitem [{\citenamefont {Refael}\ and\ \citenamefont
  {Moore}(2004)}]{MooreRefael1}%
  \BibitemOpen
  \bibfield  {author} {\bibinfo {author} {\bibfnamefont {G.}~\bibnamefont
  {Refael}}\ and\ \bibinfo {author} {\bibfnamefont {J.~E.}\ \bibnamefont
  {Moore}},\ }\href {\doibase 10.1103/PhysRevLett.93.260602} {\bibfield
  {journal} {\bibinfo  {journal} {Phys. Rev. Lett.}\ }\textbf {\bibinfo
  {volume} {93}},\ \bibinfo {pages} {260602} (\bibinfo {year}
  {2004})}\BibitemShut {NoStop}%
\bibitem [{\citenamefont {Pekker}\ \emph {et~al.}(2014)\citenamefont {Pekker},
  \citenamefont {Refael}, \citenamefont {Altman}, \citenamefont {Demler},\ and\
  \citenamefont {Oganesyan}}]{PekkerRSRGX}%
  \BibitemOpen
  \bibfield  {author} {\bibinfo {author} {\bibfnamefont {D.}~\bibnamefont
  {Pekker}}, \bibinfo {author} {\bibfnamefont {G.}~\bibnamefont {Refael}},
  \bibinfo {author} {\bibfnamefont {E.}~\bibnamefont {Altman}}, \bibinfo
  {author} {\bibfnamefont {E.}~\bibnamefont {Demler}}, \ and\ \bibinfo {author}
  {\bibfnamefont {V.}~\bibnamefont {Oganesyan}},\ }\href {\doibase
  10.1103/PhysRevX.4.011052} {\bibfield  {journal} {\bibinfo  {journal} {Phys.
  Rev. X}\ }\textbf {\bibinfo {volume} {4}},\ \bibinfo {pages} {011052}
  (\bibinfo {year} {2014})}\BibitemShut {NoStop}%
\bibitem [{\citenamefont {Vosk}\ and\ \citenamefont
  {Altman}(2013)}]{VoskAltmanPRL13}%
  \BibitemOpen
  \bibfield  {author} {\bibinfo {author} {\bibfnamefont {R.}~\bibnamefont
  {Vosk}}\ and\ \bibinfo {author} {\bibfnamefont {E.}~\bibnamefont {Altman}},\
  }\href {\doibase 10.1103/PhysRevLett.110.067204} {\bibfield  {journal}
  {\bibinfo  {journal} {Phys. Rev. Lett.}\ }\textbf {\bibinfo {volume} {110}},\
  \bibinfo {pages} {067204} (\bibinfo {year} {2013})}\BibitemShut {NoStop}%
\bibitem [{\citenamefont {Vasseur}\ \emph {et~al.}(2015)\citenamefont
  {Vasseur}, \citenamefont {Potter},\ and\ \citenamefont
  {Parameswaran}}]{QCGPRL}%
  \BibitemOpen
  \bibfield  {author} {\bibinfo {author} {\bibfnamefont {R.}~\bibnamefont
  {Vasseur}}, \bibinfo {author} {\bibfnamefont {A.~C.}\ \bibnamefont {Potter}},
  \ and\ \bibinfo {author} {\bibfnamefont {S.~A.}\ \bibnamefont
  {Parameswaran}},\ }\href {\doibase 10.1103/PhysRevLett.114.217201} {\bibfield
   {journal} {\bibinfo  {journal} {Phys. Rev. Lett.}\ }\textbf {\bibinfo
  {volume} {114}},\ \bibinfo {pages} {217201} (\bibinfo {year}
  {2015})}\BibitemShut {NoStop}%
\bibitem [{\citenamefont {Calabrese}\ and\ \citenamefont
  {Cardy}(2004)}]{1742-5468-2004-06-P06002}%
  \BibitemOpen
  \bibfield  {author} {\bibinfo {author} {\bibfnamefont {P.}~\bibnamefont
  {Calabrese}}\ and\ \bibinfo {author} {\bibfnamefont {J.}~\bibnamefont
  {Cardy}},\ }\href {\doibase 10.1088/1742-5468/2004/06/P06002} {\bibfield
  {journal} {\bibinfo  {journal} {Journal of Statistical Mechanics: Theory and
  Experiment}\ }\textbf {\bibinfo {volume} {2004}},\ \bibinfo {pages} {P06002}
  (\bibinfo {year} {2004})}\BibitemShut {NoStop}%
\bibitem [{\citenamefont {Calabrese}\ and\ \citenamefont
  {Cardy}(2009)}]{1751-8121-42-50-504005}%
  \BibitemOpen
  \bibfield  {author} {\bibinfo {author} {\bibfnamefont {P.}~\bibnamefont
  {Calabrese}}\ and\ \bibinfo {author} {\bibfnamefont {J.}~\bibnamefont
  {Cardy}},\ }\href {\doibase 10.1088/1751-8113/42/50/504005} {\bibfield
  {journal} {\bibinfo  {journal} {Journal of Physics A: Mathematical and
  Theoretical}\ }\textbf {\bibinfo {volume} {42}},\ \bibinfo {pages} {504005}
  (\bibinfo {year} {2009})}\BibitemShut {NoStop}%
\bibitem [{\citenamefont {Bransch{\"a}del}\ \emph {et~al.}(2010)\citenamefont
  {Bransch{\"a}del}, \citenamefont {Boulat}, \citenamefont {Saleur},\ and\
  \citenamefont {Schmitteckert}}]{TbRef}%
  \BibitemOpen
  \bibfield  {author} {\bibinfo {author} {\bibfnamefont {A.}~\bibnamefont
  {Bransch{\"a}del}}, \bibinfo {author} {\bibfnamefont {E.}~\bibnamefont
  {Boulat}}, \bibinfo {author} {\bibfnamefont {H.}~\bibnamefont {Saleur}}, \
  and\ \bibinfo {author} {\bibfnamefont {P.}~\bibnamefont {Schmitteckert}},\
  }\href {\doibase 10.1103/PhysRevB.82.205414} {\bibfield  {journal} {\bibinfo
  {journal} {Physical Review B}\ }\textbf {\bibinfo {volume} {82}},\ \bibinfo
  {pages} {205414} (\bibinfo {year} {2010})}\BibitemShut {NoStop}%
\bibitem [{\citenamefont {Vasseur}\ \emph {et~al.}(2014)\citenamefont
  {Vasseur}, \citenamefont {Jacobsen},\ and\ \citenamefont {Saleur}}]{VJS14}%
  \BibitemOpen
  \bibfield  {author} {\bibinfo {author} {\bibfnamefont {R.}~\bibnamefont
  {Vasseur}}, \bibinfo {author} {\bibfnamefont {J.~L.}\ \bibnamefont
  {Jacobsen}}, \ and\ \bibinfo {author} {\bibfnamefont {H.}~\bibnamefont
  {Saleur}},\ }\href {\doibase 10.1103/PhysRevLett.112.106601} {\bibfield
  {journal} {\bibinfo  {journal} {Physical review letters}\ }\textbf {\bibinfo
  {volume} {112}},\ \bibinfo {pages} {106601} (\bibinfo {year}
  {2014})}\BibitemShut {NoStop}%
\bibitem [{\citenamefont {Freton}\ \emph {et~al.}(2013)\citenamefont {Freton},
  \citenamefont {Boulat},\ and\ \citenamefont {Saleur}}]{EEIRLM}%
  \BibitemOpen
  \bibfield  {author} {\bibinfo {author} {\bibfnamefont {L.}~\bibnamefont
  {Freton}}, \bibinfo {author} {\bibfnamefont {E.}~\bibnamefont {Boulat}}, \
  and\ \bibinfo {author} {\bibfnamefont {H.}~\bibnamefont {Saleur}},\ }\href
  {\doibase 10.1016/j.nuclphysb.2013.05.015} {\bibfield  {journal} {\bibinfo
  {journal} {Nuclear Physics B}\ }\textbf {\bibinfo {volume} {874}},\ \bibinfo
  {pages} {279} (\bibinfo {year} {2013})}\BibitemShut {NoStop}%
\bibitem [{Sup()}]{SupMat}%
  \BibitemOpen
  \href@noop {} {\bibinfo  {journal} {See supplemental material}\ }\BibitemShut
  {NoStop}%
\bibitem [{\citenamefont {Ma}\ \emph {et~al.}(1979)\citenamefont {Ma},
  \citenamefont {Dasgupta},\ and\ \citenamefont {Hu}}]{PhysRevLett.43.1434}%
  \BibitemOpen
\bibfield  {journal} {  }\bibfield  {author} {\bibinfo {author} {\bibfnamefont
  {S.-k.}\ \bibnamefont {Ma}}, \bibinfo {author} {\bibfnamefont
  {C.}~\bibnamefont {Dasgupta}}, \ and\ \bibinfo {author} {\bibfnamefont
  {C.-k.}\ \bibnamefont {Hu}},\ }\href {\doibase 10.1103/PhysRevLett.43.1434}
  {\bibfield  {journal} {\bibinfo  {journal} {Phys. Rev. Lett.}\ }\textbf
  {\bibinfo {volume} {43}},\ \bibinfo {pages} {1434} (\bibinfo {year}
  {1979})}\BibitemShut {NoStop}%
\bibitem [{\citenamefont {Dasgupta}\ and\ \citenamefont
  {Ma}(1980)}]{PhysRevB.22.1305}%
  \BibitemOpen
  \bibfield  {author} {\bibinfo {author} {\bibfnamefont {C.}~\bibnamefont
  {Dasgupta}}\ and\ \bibinfo {author} {\bibfnamefont {S.-k.}\ \bibnamefont
  {Ma}},\ }\href {\doibase 10.1103/PhysRevB.22.1305} {\bibfield  {journal}
  {\bibinfo  {journal} {Phys. Rev. B}\ }\textbf {\bibinfo {volume} {22}},\
  \bibinfo {pages} {1305} (\bibinfo {year} {1980})}\BibitemShut {NoStop}%
\bibitem [{\citenamefont {Fisher}\ and\ \citenamefont
  {Young}(1998)}]{FisherYoung}%
  \BibitemOpen
  \bibfield  {author} {\bibinfo {author} {\bibfnamefont {D.~S.}\ \bibnamefont
  {Fisher}}\ and\ \bibinfo {author} {\bibfnamefont {A.~P.}\ \bibnamefont
  {Young}},\ }\href {\doibase 10.1103/PhysRevB.58.9131} {\bibfield  {journal}
  {\bibinfo  {journal} {Phys. Rev. B}\ }\textbf {\bibinfo {volume} {58}},\
  \bibinfo {pages} {9131} (\bibinfo {year} {1998})}\BibitemShut {NoStop}%
\bibitem [{\citenamefont {Fagotti}\ \emph {et~al.}(2011)\citenamefont
  {Fagotti}, \citenamefont {Calabrese},\ and\ \citenamefont
  {Moore}}]{CalabreseMoore}%
  \BibitemOpen
  \bibfield  {author} {\bibinfo {author} {\bibfnamefont {M.}~\bibnamefont
  {Fagotti}}, \bibinfo {author} {\bibfnamefont {P.}~\bibnamefont {Calabrese}},
  \ and\ \bibinfo {author} {\bibfnamefont {J.~E.}\ \bibnamefont {Moore}},\
  }\href {\doibase 10.1103/PhysRevB.83.045110} {\bibfield  {journal} {\bibinfo
  {journal} {Phys. Rev. B}\ }\textbf {\bibinfo {volume} {83}},\ \bibinfo
  {pages} {045110} (\bibinfo {year} {2011})}\BibitemShut {NoStop}%
\bibitem [{\citenamefont {{Devakul}}\ \emph {et~al.}(2016)\citenamefont
  {{Devakul}}, \citenamefont {{Majumdar}},\ and\ \citenamefont
  {{Huse}}}]{2016arXiv161202012D}%
  \BibitemOpen
  \bibfield  {author} {\bibinfo {author} {\bibfnamefont {T.}~\bibnamefont
  {{Devakul}}}, \bibinfo {author} {\bibfnamefont {S.~N.}\ \bibnamefont
  {{Majumdar}}}, \ and\ \bibinfo {author} {\bibfnamefont {D.~A.}\ \bibnamefont
  {{Huse}}},\ }\href@noop {} {\bibfield  {journal} {\bibinfo  {journal} {ArXiv
  e-prints}\ } (\bibinfo {year} {2016})},\ \Eprint
  {http://arxiv.org/abs/1612.02012} {arXiv:1612.02012 [cond-mat.stat-mech]}
  \BibitemShut {NoStop}%
\bibitem [{\citenamefont {Peschel}(2005)}]{0305-4470-38-20-002}%
  \BibitemOpen
  \bibfield  {author} {\bibinfo {author} {\bibfnamefont {I.}~\bibnamefont
  {Peschel}},\ }\href {\doibase 10.1088/0305-4470/38/20/002} {\bibfield
  {journal} {\bibinfo  {journal} {Journal of Physics A: Mathematical and
  General}\ }\textbf {\bibinfo {volume} {38}},\ \bibinfo {pages} {4327}
  (\bibinfo {year} {2005})}\BibitemShut {NoStop}%
\bibitem [{\citenamefont {Peschel}\ and\ \citenamefont
  {Eisler}(2009)}]{Peschel}%
  \BibitemOpen
  \bibfield  {author} {\bibinfo {author} {\bibfnamefont {I.}~\bibnamefont
  {Peschel}}\ and\ \bibinfo {author} {\bibfnamefont {V.}~\bibnamefont
  {Eisler}},\ }\href {http://stacks.iop.org/1751-8121/42/i=50/a=504003}
  {\bibfield  {journal} {\bibinfo  {journal} {Journal of Physics A:
  Mathematical and Theoretical}\ }\textbf {\bibinfo {volume} {42}},\ \bibinfo
  {pages} {504003} (\bibinfo {year} {2009})}\BibitemShut {NoStop}%
\bibitem [{\citenamefont {Laflorencie}(2005)}]{Laflorencie}%
  \BibitemOpen
  \bibfield  {author} {\bibinfo {author} {\bibfnamefont {N.}~\bibnamefont
  {Laflorencie}},\ }\href {\doibase 10.1103/PhysRevB.72.140408} {\bibfield
  {journal} {\bibinfo  {journal} {Phys. Rev. B}\ }\textbf {\bibinfo {volume}
  {72}},\ \bibinfo {pages} {140408} (\bibinfo {year} {2005})}\BibitemShut
  {NoStop}%
\bibitem [{\citenamefont {Vasseur}\ and\ \citenamefont
  {Moore}(2015)}]{MooreVasseur}%
  \BibitemOpen
  \bibfield  {author} {\bibinfo {author} {\bibfnamefont {R.}~\bibnamefont
  {Vasseur}}\ and\ \bibinfo {author} {\bibfnamefont {J.~E.}\ \bibnamefont
  {Moore}},\ }\href {\doibase 10.1103/PhysRevB.92.054203} {\bibfield  {journal}
  {\bibinfo  {journal} {Phys. Rev. B}\ }\textbf {\bibinfo {volume} {92}},\
  \bibinfo {pages} {054203} (\bibinfo {year} {2015})}\BibitemShut {NoStop}%
\bibitem [{\citenamefont {Westerberg}\ \emph {et~al.}(1995)\citenamefont
  {Westerberg}, \citenamefont {Furusaki}, \citenamefont {Sigrist},\ and\
  \citenamefont {Lee}}]{PhysRevLett.75.4302}%
  \BibitemOpen
  \bibfield  {author} {\bibinfo {author} {\bibfnamefont {E.}~\bibnamefont
  {Westerberg}}, \bibinfo {author} {\bibfnamefont {A.}~\bibnamefont
  {Furusaki}}, \bibinfo {author} {\bibfnamefont {M.}~\bibnamefont {Sigrist}}, \
  and\ \bibinfo {author} {\bibfnamefont {P.~A.}\ \bibnamefont {Lee}},\ }\href
  {\doibase 10.1103/PhysRevLett.75.4302} {\bibfield  {journal} {\bibinfo
  {journal} {Phys. Rev. Lett.}\ }\textbf {\bibinfo {volume} {75}},\ \bibinfo
  {pages} {4302} (\bibinfo {year} {1995})}\BibitemShut {NoStop}%
\bibitem [{\citenamefont {Igl{\'o}i}\ and\ \citenamefont
  {Monthus}(2005)}]{Monthus}%
  \BibitemOpen
  \bibfield  {author} {\bibinfo {author} {\bibfnamefont {F.}~\bibnamefont
  {Igl{\'o}i}}\ and\ \bibinfo {author} {\bibfnamefont {C.}~\bibnamefont
  {Monthus}},\ }\href {\doibase
  http://dx.doi.org/10.1016/j.physrep.2005.02.006} {\bibfield  {journal}
  {\bibinfo  {journal} {Physics Reports}\ }\textbf {\bibinfo {volume} {412}},\
  \bibinfo {pages} {277 } (\bibinfo {year} {2005})}\BibitemShut {NoStop}%
\bibitem [{\citenamefont {Filyov}\ \emph {et~al.}(1981)\citenamefont {Filyov},
  \citenamefont {Tzvelik},\ and\ \citenamefont {Wiegmann}}]{FTW80}%
  \BibitemOpen
  \bibfield  {author} {\bibinfo {author} {\bibfnamefont {V.}~\bibnamefont
  {Filyov}}, \bibinfo {author} {\bibfnamefont {A.}~\bibnamefont {Tzvelik}}, \
  and\ \bibinfo {author} {\bibfnamefont {P.}~\bibnamefont {Wiegmann}},\ }\href
  {\doibase 10.1016/0375-9601(81)90055-4} {\bibfield  {journal} {\bibinfo
  {journal} {Physics Letters A}\ }\textbf {\bibinfo {volume} {81}},\ \bibinfo
  {pages} {175} (\bibinfo {year} {1981})}\BibitemShut {NoStop}%
\end{thebibliography}%

\bigskip


%

\appendix
\newpage

\onecolumngrid
{\center \large \bf Healing of Defects in Random Antiferromagnetic Spin Chains: Supplemental Material}

\section{Probability of decimating the weak link }
  
We recall from the main text that  the probability distribution for the weak link is initially
\begin{equation}
Q_{\Gamma_0} (\beta) = \frac{1}{\Gamma_0} {\rm e}^{-\frac{\beta + \ln \lambda}{\Gamma_0}} \theta(\beta + \ln \lambda),
\label{eqInitialCond}
\end{equation}
with $\Gamma_0$ the initial disorder strength, since initially the weak link has $\beta = \ln \frac{\Omega_0}{\lambda J_0} > \ln \lambda^{-1}$. The probability $p_\Gamma$ for this weak link not to be decimated at scale $\Gamma > \Gamma_0$ in the RG is given by
\begin{equation}
\frac{d p_\Gamma}{d \Gamma} = - Q_{\Gamma}(0), \ \text{and} \ \int_{0}^\infty Q_{\Gamma}(\beta) d \beta = p_\Gamma,
\end{equation}
where $Q_{\Gamma}(\beta)$ satisfies the flow equation 
\begin{equation}
\frac{\partial Q_\Gamma}{\partial \Gamma} = \frac{\partial Q_\Gamma}{\partial \beta} + 2 P_\Gamma(0) \left(P_\Gamma \star Q_\Gamma - Q_\Gamma \right).
\label{eqPDE}
\end{equation}
Computing $p_\Gamma$ therefore amounts to solving this partial differential equation with the initial condition~\eqref{eqInitialCond}. Some simple solutions of this equation of the form $Q_\Gamma (\beta)= \left(a_\Gamma + \frac{\beta}{\Gamma} b_\Gamma \right) P_\Gamma(\beta)$  were found in~\cite{MooreRefael1}, where $a_\Gamma$ and $b_\Gamma$  ordinary differential equations. However, our initial condition~\eqref{eqInitialCond} is more complicated and is not of this form because of the Heaviside function. To proceed, we introduce the Laplace transform 
\begin{equation}
\hat{Q}_\Gamma (s) = \int_0^\infty {\rm e}^{- s \beta} Q_\Gamma (\beta) d \beta,
\end{equation}
which satisfies
\begin{equation}
\frac{\partial \hat{Q}_\Gamma (s)}{\partial \Gamma} + \frac{1 - s \Gamma}{1+s \Gamma} s \hat{Q}_\Gamma (s) = - Q_\Gamma(\beta=0) = \frac{d p_\Gamma}{d \Gamma}.
\end{equation}
This gives
\begin{equation}
\hat{Q}_\Gamma (s) = \left( \frac{1+s \Gamma_0}{1+s \Gamma} \right)^2 {\rm e}^{s(\Gamma - \Gamma_0)} \left[\hat{Q}_{\Gamma_0} (s) + \int_{\Gamma_0}^\Gamma \frac{d p_{\Gamma^\prime}}{d {\Gamma^\prime}} \left( \frac{1+s \Gamma^\prime}{1+s \Gamma_0} \right)^2 {\rm e}^{-s(\Gamma^\prime - \Gamma_0)} d {\Gamma^\prime} \right],
\label{eqLaplace}
\end{equation}
with $\hat{Q}_{\Gamma_0} (s) = \lambda^s/(1+\Gamma_0 s)$.
Note that we have $p_\Gamma= \hat{Q}_\Gamma (s=0) $ as we should. Now the unknown $\frac{d p_{\Gamma}}{d {\Gamma}}$ can be obtained self-consistently as the inverse Laplace transform
\begin{equation}
\frac{d p_{\Gamma}}{d {\Gamma}} = - Q_\Gamma(\beta=0) = \int_{c-i \infty}^{c+i \infty} \frac{d s}{2 \pi i} \hat{Q}_\Gamma (s).
\end{equation}
Taking the inverse Laplace transform of eq.~\eqref{eqLaplace}, we find that $\frac{d p_{\Gamma}}{d {\Gamma}} $ satisfies the integral equation
\begin{align}
-\frac{d p_{\Gamma}}{d {\Gamma}} &= \frac{1}{\Gamma} {\rm e}^{\Gamma_0/\Gamma-1}{\rm e}^{-\log \lambda/\Gamma} \theta(\Gamma-\Gamma_0 + \log \lambda) \left[ 1 - \frac{\Gamma_0}{\Gamma} + \left( \frac{\Gamma_0}{\Gamma} \right)^2 + \left( 1 - \frac{\Gamma_0}{\Gamma}\right) \frac{\log \lambda}{\Gamma}\right] \notag \\
& +   \int_{\Gamma_0}^\Gamma d {\Gamma^\prime} \frac{d p_{\Gamma^\prime}}{d {\Gamma^\prime}} {\rm e}^{\Gamma^\prime/\Gamma-1} \left( 1 - \frac{\Gamma^\prime}{\Gamma}\right)\left( 1 + \left(\frac{\Gamma^\prime}{\Gamma} \right)^2\right).
\end{align}
Although this looks quite complicated, one can check   that for $\lambda=1$, this equation is satisfied by the exact solution for $\frac{d p_{\Gamma}}{d {\Gamma}} $ that can be computed very simply in that case. We now consider the universal regime $\Gamma \to \infty$, $\lambda \to 0$ with $x = \frac{\Gamma}{-\ln \lambda}$ fixed: this amounts to sending $\Gamma_0 \to 0$. In this regime, we find
\begin{align}
-x \frac{d p}{d x} &= {\rm e}^{1/x-1} \left( 1 - \frac{1}{x}\right) \theta(x-1) + \int_0^x dx^\prime \frac{d p}{d x^\prime} {\rm e}^{x^\prime/x-1}   \left( 1 - \frac{x^\prime}{x}\right)\left( 1 + \left(\frac{x^\prime}{x} \right)^2\right).
\label{eqIntEq}
\end{align}
Using the fact that $p(x)$ should decay as $x^{- \frac{3-\sqrt{5}}{2}}$  and with a little bit of guess work, we find that the solution of this equation is
\begin{equation}
p(x) = \frac{1 - \sqrt{5}}{5+\sqrt{5}} x^{- \frac{3+\sqrt{5}}{2}} + \frac{4 + 2 \sqrt{5}}{5+\sqrt{5}}  x^{- \frac{3-\sqrt{5}}{2}}, \ x>1,
\label{eqPx}
\end{equation}
and $p(x)=1$ for $x \leq 1$. This is in agreement with the result found in the main text using a different approach (see below). The probability of not decimating the weak link studied in the main text then reads
\begin{equation}
P_0\left( \frac{L}{L^\star}\right) = \frac{4}{\pi} \sum_{n=0}^\infty \frac{(-1)^n}{2n+1} {\rm e}^{- \frac{\pi^2}{4} \frac{L}{L^\star} (2n+1)^2} + 2 \pi \frac{L}{L^\star} \sum_{n=0}^\infty (-1)^n (2 n+1) \int_{1}^\infty \frac{dx}{x^3} p(x) {\rm e}^{- \frac{\pi^2}{4 x^2} \frac{L}{L^\star} (2n+1)^2}.
\end{equation}

\section{Universal entanglement crossover}

Let us now go back to the entanglement calculation. Following Refael and Moore~\cite{MooreRefael1}, we compute the average RG time $\ell = \overline{\ln \Gamma}$ corresponding to the decimation of the weak link. This quantity is given by 
\begin{equation}
\overline{\ln \frac{\Gamma}{-\ln \lambda}} = \int_0^\infty dx \frac{dp}{dx} \ln x = 3, 
\end{equation}
 where we used eq.~\eqref{eqPx}. Remarkably, this is the same value that Refael and Moore~\cite{MooreRefael1} find for the average RG time separating decimations that increase  the entanglement entropy by $\ln 2$: in a setup with $\lambda=1$, if the central link is decimated at some scale $\Gamma$, it is typically much weaker than the other bonds in the chain because of the renormalization factor and it is typically decimated again at the scale $\Gamma^\prime$ given by $\ln \frac{\Gamma^\prime}{\Gamma}=3$. This suggests the following very simple picture: with or without the weak link, the average RG time between decimations increasing the entanglement entropy by $\ln 2$ is given by $\ell = 3$, but the effect of the weak link is merely to shift the origin of time, since the first decimation typically occurs when $\overline{\ln \frac{\Gamma}{-\ln \lambda}} =3$, with in particular $\Gamma> - \ln \lambda$. 
 
Actually, it turns out that~\eqref{eqPx} is exactly the expression that Refael and Moore found for the probability of a bond decimated at scale $\Gamma_0$, with initial probability distribution 
\begin{equation}
Q_{\Gamma_0} (\beta) = \frac{\beta}{\Gamma_0^2} {\rm e}^{-\beta/\Gamma_0},
\label{eqRM}
\end{equation}
to not have  been decimated again at $x=\Gamma/\Gamma_0$. This suggests a much easier way to derive~\eqref{eqPx} (used in the main text), even if our initial condition~\eqref{eqInitialCond} is completely different. The effect of the weak link is to shift the origin of time (which makes sense since the bond cannot be decimated if $\Gamma< - \ln \lambda$), and the probability distribution of the weak link at that scale $\Gamma=-\ln \lambda$ should coincide with~\eqref{eqRM} where $\Gamma_0$ is replaced by $-\ln \lambda$ (the effect of the renormalization due to the decimations of its neighbors is ``as if'' the weak bond had just been decimated). This can be checked explicitly, as $dp_\Gamma/d\Gamma=0$ if $\Gamma \leq - \ln \lambda$ so that the inverse Laplace transform of~\eqref{eqLaplace} for $\Gamma = -\ln \lambda \gg \Gamma_0$  reads
\begin{equation}
Q_{\Gamma=-\ln \lambda} (\beta) =\int_{c -i \infty}^{c+i \infty} \frac{ds}{ 2 \pi i} {\rm e}^{s \beta} \frac{1}{(1-s \ln \lambda)^2} =  \frac{\beta}{(\ln \lambda)^2} {\rm e}^{\beta/\ln \lambda},
\label{eqNewInitialCondition}
\end{equation}
which indeed coincides with~\eqref{eqRM} with $\Gamma_0$ replaced by $-\ln \lambda$ as claimed. This means that instead of solving by pure guess work the integral equation~\eqref{eqIntEq} to find $p(x)$, we can actually solve the equation~\eqref{eqPDE} with the much simpler ``initial condition''~\eqref{eqNewInitialCondition} using the results of Refael and Moore.  

Not only does this observation allow us to obtain the result~\eqref{eqPx} in a much simpler way, it gives us a clear way to compute entanglement. The physical picture is clear: the average RG times between decimations is always $3$, but the effect of the weak link is to shift the origin of time. This gives for the bipartite entanglement entropy with periodic BCs
\begin{equation}
S = \int_{\Gamma_0}^\infty    \frac{d \Gamma}{\Gamma} g \left( \frac{\Gamma}{\sqrt{L}} \right) \ln 2 \left[ \int_{\Gamma_0}^\Gamma  \frac{d \Gamma^\prime/\Gamma^\prime}{3} +  \int_{\Gamma_0}^\Gamma \frac{ d \Gamma^\prime/\Gamma^\prime}{3} \theta(\Gamma^\prime + \ln \lambda)\right],
\end{equation}
up to non-universal contributions. Introducing $u=\Gamma/\sqrt{L}$ and $x=\Gamma^\prime/(-\ln \lambda)$, this yields
\begin{equation}
S = \int_{\Gamma_0/\sqrt{L}}^\infty    \frac{d u}{u} g(u)  \frac{\ln 2}{3}  \ln \left( \frac{u \sqrt{L}}{\Gamma_0} \right) + \int_{\Gamma_0/\sqrt{L}}^\infty    \frac{d u}{u} g(u)  \frac{\ln 2}{3} \int_{\Gamma_0/(-\ln \lambda)}^{u \sqrt{L/L^\star}}\frac{dx}{x} \theta(x-1). 
\end{equation}
If we now take the scaling limit $\Gamma \to \infty$, $\lambda \to 0$ with $x = \frac{\Gamma}{-\ln \lambda}$ fixed (with $\Gamma_0/\ln \lambda^{-1} \to 0$ and $\Gamma_0/\sqrt{L} \to 0$), and consider the derivative with respect to $\ln L$ to get rid of the non-universal terms, we get
\begin{equation}
\frac{\partial S}{ \partial \ln L} = \frac{\ln 2}{6} + \int_{0}^\infty    \frac{d u}{u} g(u)  \frac{\ln 2}{3} \frac{\partial (u \sqrt{L/L^\star})}{\partial \ln L} \frac{1}{u \sqrt{L/L^\star}} \theta(u \sqrt{L/L^\star}-1),
\end{equation}
where we recall that $\int_{0}^\infty    \frac{d u}{u} g(u) =1$. This yields
\begin{equation}
\frac{\partial S}{ \partial \ln L} = \frac{\ln 2}{6} \left( 1 + \int_{\sqrt{L^\star/L}}^\infty \frac{d u}{u} g(u)  \right).
\end{equation}
Plugging in the expression for $g$, we finally get
\begin{equation}
\frac{\partial S}{ \partial \ln L} = \frac{\ln 2}{3} \left( 1  - \frac{2}{\pi} \sum_{n=0}^\infty \frac{(-1)^n}{2n+1} {\rm e}^{-\pi^2 (2n+1)^2 L /4 L^\star} \right),
\label{eqSapprox}
\end{equation}
as claimed in the main text.

\section{Entanglement spectrum and distribution of entanglement across the weak link}

Using the picture above and the results of~\cite{CalabreseMoore}, we can immediately compute the scaling form of the generating function of the number of singlets $n$ crossing the entanglement cuts (one of which being the weak link), a quantity directly related to the entanglement spectrum. We find
\begin{equation}
\langle {\rm e }^{n t}\rangle =  \int_0^\infty d \Gamma  \frac{1}{\Gamma} g \left( \frac{\Gamma}{\sqrt{L}} \right) \left[ \theta \left(1- \frac{\Gamma}{-\ln \lambda} \right) + \theta \left( \frac{\Gamma}{-\ln \lambda} -1 \right) \left( \alpha_t \left(\frac{\Gamma}{-\ln \lambda} \right)^{- \frac{3-\sqrt{5+4 {\rm e}^t}}{2}}  + \beta_t \left(\frac{\Gamma}{-\ln \lambda} \right)^{- \frac{3+\sqrt{5+4 {\rm e}^t}}{2}} \right)   \right],
\end{equation}
where the coefficients $\alpha_t$ and $\beta_t$ are given in Ref.~\cite{CalabreseMoore}. Let us introduce the function
\begin{equation}
I_\alpha (x) = \int_{x}^\infty \frac{du}{u} g(u) u^{-\alpha} = \left( \frac{2}{\pi} \right)^{1+\alpha} \sum_{n=0}^{\infty} \frac{(-1)^n}{(1+2n)^{\alpha+ 1}} \left( \alpha \Gamma_{\frac{\alpha}{2}} - 2 \Gamma_{1+\frac{\alpha}{2}, \frac{x \pi^2}{4 }(1+2n)^2}\right),
\end{equation}
so that 
\begin{equation}
\langle {\rm e }^{n t}\rangle =  1 - I_0 \left( \frac{L}{L_\star}\right) + \alpha_t \left( \frac{L}{L_\star}\right) ^{- \frac{3-\sqrt{5+4 {\rm e}^t}}{4}} I_{\frac{3-\sqrt{5+4 {\rm e}^t}}{2}}  \left( \frac{L}{L_\star}\right) + \beta_t \left( \frac{L}{L_\star}\right) ^{- \frac{3+\sqrt{5+4 {\rm e}^t}}{4}} I_{\frac{3+\sqrt{5+4 {\rm e}^t}}{2}}  \left( \frac{L}{L_\star}\right).
\end{equation}
The limit $ t \to -\infty$ gives $P_0 = \lim_{t \to -\infty} \langle {\rm e }^{n t}\rangle$, the probability for the weak link not to be decimated for a system of size $L$, in agreement with the formula given in the main text. The entanglement entropy also follows from taking a derivative with respect to $t$ and then letting $t \to 0$ to obtain the mean number of singlets crossing the entanglement cuts $\langle n \rangle $. Interestingly, the resulting formula contains correction compared to eq.~\eqref{eqSapprox}. This is because in deriving~\eqref{eqSapprox}, we used the fact that the number of singlets crossing a generic link in the system scales as $n \sim \frac{1}{3} \ln \frac{\Gamma}{\Gamma0}$ in terms of the RG scale $\Gamma$~\cite{MooreRefael1}. Ref.~\cite{CalabreseMoore} predicts some corrections to this formula, with $n \sim \frac{1}{3} \ln\frac{\Gamma}{\Gamma0} + \frac{1}{9} \left( (\Gamma_0/\Gamma)^{3} - 1\right) +\dots$ Whereas these subleading corrections can essentially be ignored for generic links in the scaling limit where $\Gamma  \gg \Gamma_0$, they are non-negligible in our case since $\Gamma_0$ is effectively replaced by the crossover scale $-\ln \lambda$ when computing the entanglement across the weak link. By implementing the RSRG procedure numerically, we found that these corrections to~\eqref{eqSapprox} improve the agreement with the numerical results (see main text). It is however unclear to us whether the results of~\cite{CalabreseMoore} exhaust all the important subleading corrections, and to what extent these corrections are universal (as they should be to enter the universal scaling function for the entanglement entropy).

\end{document}